# Importance of suppression and mitigation measures in managing COVID-19 outbreaks

Michael E. Hochberg

ISEM, University of Montpellier, CNRS, IRD, EPHE, Montpellier, France

Santa Fe Institute, Santa Fe, New Mexico, USA

michael.hochberg@umontpellier.fr

## Abstract

I employ a simple mathematical model of an epidemic process to evaluate how four basic quantities: the reproduction number ($\mathcal{R}$), the numbers of sensitive (S) and infectious individuals (I), and total community size (N) affect strategies to control COVID-19. Numerical simulations show that strict suppression measures at the beginning of an epidemic can create low infectious numbers, which thereafter can be managed by mitigation measures over longer periods to flatten the epidemic curve. The stronger the suppression measure, the faster it achieves the low numbers of infections that are conducive to subsequent management. Our results on short-term strategies point to either a two-step control strategy, following failed mitigation, that begins with suppression of the reproduction number, $\mathcal{R}_C$, below 1.0, followed by renewed mitigation measures that manage the epidemic by maintaining $\mathcal{R}_C$ at approximately 1.0, or should suppression not be feasible, the progressive lowering of the effective reproductive number, $\mathcal{R}_{Ceff} \approx \mathcal{R}_C$ S/N, below 1.0. The objectives of the full sequence of measures observed in a number of countries, and likely to see in the longer term, can be symbolically represented as: $\mathcal{R}_0$ → $\mathcal{R}_C<\mathcal{R}_0$ → $\mathcal{R}_C<<1.0$ → $\mathcal{R}_C\approx 1.0$ → $\mathcal{R}_{Ceff}<1.0$. We discuss the predictions of this analysis and how it fits into longer-term sequences of measures, including using the herd immunity concept to 'leverage' acquired immunity.

**Introduction**

The COVID-19 pandemic is a major global threat. Spread of the SARS-CoV-2 virus began in China in late 2019, with 41 cases recorded according to the WHO as of January 11/12, 2020, attaining almost 4 million confirmed cases worldwide as of May 6th (*1*). By its highly transmissible and virulent nature, COVID-19 is putting considerable strain on health services, meaning that increasing numbers of patients in the most afflicted countries cannot be adequately cared for, which will likely further exacerbate disease morbidity and mortality.

Research groups have mobilized to collect and analyze molecular (*2*) and epidemiological (*3-5*) data, and employ statistical and mathematical models to simulate regional and national outbreaks and the global pandemic, and evaluate possible control measures (e.g., *6-14*).

Particularly important in this effort is the projection of how different strategies will affect outbreaks. Conducting such studies without delay is crucial, both because most countries are in early outbreak stages and thus open to management options, and since some nations are days or even weeks behind in their epidemics compared to others. The latter property is important, since the lockstep nature of COVID-19 epidemic trends mean that nations can 'peer into the future' to predict how their own outbreaks could unfold. This information and the efficacy of control strategies already adopted by other 'future' countries can be instrumental in giving the time needed to plan and logistically organize effective measures.

Here, we employ a simple epidemiological model to elucidate some of the basic parameters and processes that arbitrate short-term control measure outcomes. The model's intuitive results emphasize the necessity to adopt one or both of two strategies. 'Suppression measures' are engaged early and decisively to lower the reproduction number, $\mathcal{R}$, below 1.0 and as close as possible to 0.0. When successful, this results in low, manageable numbers of infections (a low 'set-point'), and can then be followed by a second strategy: 'Mitigation measures' that continue to flatten the epidemic curve by maintaining the reproduction number to approximately 1.0. However, strategies need not follow this sequence. For those countries reacting very early in an outbreak, mitigation measures alone may be sufficient. For those that are late and/or unable to employ radical suppression measures, 'progressive mitigation' so as to lower $\mathcal{R}$ below 1.0 may be the only alternative.

The objective of the present study is to emphasize how strategic lowering of the reproduction number is central to a rational management plan to minimize the impacts of COVID-19. Because local outbreaks are producing large numbers of recovered cases, this opens the possibility that any acquired immunity can be 'leveraged' to reduce the intensity of future policies.

## Model

We employ a modified SEIR model of Susceptible (*S*) → Exposed (*E*) → Infectious (*I*) → Removed (*R*) states (*15*). The ordinary differential equations take the form:

$$\frac{dS}{dt} = -\beta \frac{IS}{N}$$

$$\frac{dE}{dt} = \beta \frac{IS}{N} - \frac{1}{T_{inf}} E$$

$$\frac{dI}{dt} = \frac{1}{T_{inf}} E - \gamma I$$

$$\frac{dR}{dt} = \gamma I,$$

where *N* is a constant equal to *S+E+I+R*, $\beta$ is the transmission parameter, $T_{inf}$ is the infectious period, $\gamma$ is the rate of removal into different subclasses of *R*. Specifically, *R* is composed of minor cases $C_M$, severe cases at home $C_S$, severe cases in hospital $C_{SH}$, and fatalities $C_F$, given by

$$\frac{dC_M}{dt} = P_M \gamma I - \frac{C_M}{D_{RM}},$$

$$\frac{dC_S}{dt} = P_S \gamma I - \frac{C_S}{D_{HL}},$$

$$\frac{dC_{SH}}{dt} = \frac{C_S}{D_{HL}} - \frac{C_{SH}}{D_{RS}},$$

$$\frac{dC_F}{dt} = P_F \gamma I - \frac{C_F}{D_F}.$$

Here, $P_M$, $P_S$, and $P_F$ are, respectively, the probabilities of a case being mild, severe and resulting in death ($P_M + P_S + P_F = 1$). $D_{RM}$, $D_{RS}$, $D_{HL}$, are days to recovery for mild and severe cases, time of hospitalization, and $D_F$ is the time from end of incubation to death ($T_{itod}$) minus duration of infectiousness ($T_{inf}$).

An outbreak occurs if the basic reproduction number, $\mathcal{R}_0 = \beta/\gamma > 1.0$. The impact of control measures is easily understood by their impact on $\mathcal{R}_0$, and in the presentation below, we refer to these effects by reductions in $\mathcal{R}_0$, yielding the modified constant value, $\mathcal{R}_C$.

# Numerical methods

We employ this general model for COVID-19 to explore how some of its central properties could affect outbreak control efforts. The over-simplicity of the model means that it should not be used to make precise predictions for actual epidemic management situations, but rather serve as a conceptual tool that can serve as a first step towards more realistic analyses for specific scenarios or situations.

Specifically, we focus on how two generic types of objective – suppression and mitigation – affect epidemiological and clinical parameters. We expect that, all else being equal, suppression will reduce $\mathcal{R}_0$ more than would mitigation. However, their impacts on $\mathcal{R}_0$ are expected to differ from locality to locality depending on the details of (and adherence to) the measures deployed. This argues for examining a range of $\mathcal{R}_C$ for each of the two types of objective, and recognizing that mitigation measures in one locality (e.g., country) could be more effective at lowering $\mathcal{R}_0$ than suppression measures in another.

Epidemic management strategies were investigated using the Epidemic Calculator package (*16*) (Supplementary Material). This platform is rich in possibilities for varying key parameters such as the reproduction number ($\mathcal{R}$) and temporal scales of infection ($T_{inf}$ and $T_{inc}$), as well as initial sub-population of infections individuals ($I_0$) and the total population size ($N=S_0+I_0$). The platform also permits the user to experiment with different "clinical" parameters, including hospitalization rate, case fatality rate, and recovery time for mild cases.

The results presented below are based on the parameter values provided on the website simulator page (**Table 1**). Given the recent emergence of COVID-19, these parameter values should be viewed as preliminary and possibly inaccurate, since for example, they may be based on limited data or be time- or location-specific. This reinforces the above call for caution in interpreting the findings presented here, and in using the precise model output for any specific management actions.

| Parameter | Value |
|---|---|
| Basic Reproduction Number ($\mathcal{R}_0$) | 2.5, 3.0 |
| Removal rate ($\gamma$) | 1.0/5.2 days$^{-1}$ |
| Duration patient is infectious ($T_{inf}$) | 2.9 days |
| Case fatality rate ($P_F$) | 0.02 |

| Time from end of incubation to death ($T_{itod}$) | 32 days |
|---|---|
| Recovery time for severe cases ($D_{RS}$) | 28.6 days |
| Recovery time for mild cases ($D_{RM}$) | 11.1 days |
| Hospitalization rate ($P_S$) | 0.2 |
| Time to hospitalization ($D_{HL}$) | 5 days |

**Table 1.** Parameters and their baseline values employed in this study. $\mathcal{R}_0$ values investigated are based on (*3,6*). See (*16*) for details on other parameters.

The Epidemic Calculator package, although very flexible, has some limitations in its use for scientific study. First, the accuracy of the simulator output is untested with respect to analytical results, independent computational studies, and other SEIR platforms. The only test conducted here did verifiy that the simulated equilibrium fraction of susceptible individuals ($S_\infty$) followed the predicted relation: $-\ln(S_\infty/S_0) = \mathcal{R}(1-S_\infty/S_0)$ (*15*). The results presented below – even if consistent with intuition – nevertheless need to be viewed as preliminary and contingent on future testing. Second, precision in the intervals for input parameters and platform output were not always to the last decimal places, and for large numbers, such as the total community size N, a limited number of choices were available. Therefore, for example, there was no choice for exactly N=70 million, and as such the next highest option (70,420,854) was employed. The same was true for simulations with the lower population size of N=70K (70,263 was used). Moreover, the data presented below (i.e., y-axis data point readings) were in some cases closest interpolations of closely neighboring values that resulted in the simulation target. Varying the input and reading rules to neighboring values was found to have negligible effects on the trends reported, and did not change the main conclusions of this study.

## Results

We first explored how the trigger number of infectious cases (I) for suppression measures to be engaged, affected the critical level of $\mathcal{R}_C$ necessary to keep infectious cases at or below a set-point of 100 after 60 days. We chose 60 days because it is the approximate period that areas of China (as the first affected country), decided to enter lockdown.

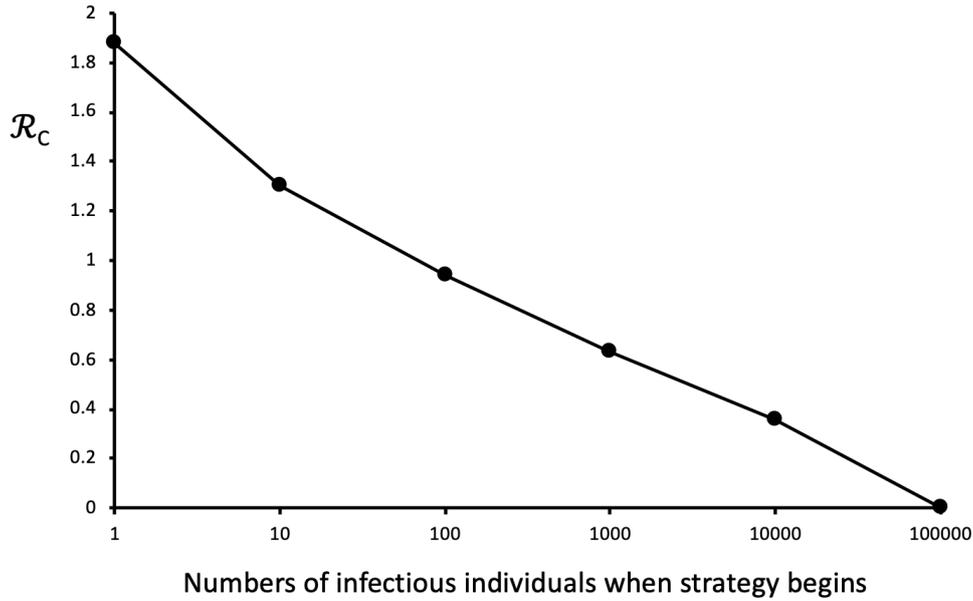

**Figure 1.** Suppression levels required to meet objectives, given different starting conditions. Simulations begin with no measure imposed ($R_0$=2.5) and 1 infectious individual in a population of *c.* 70 million unexposed individuals. The suppression strategy starts when the number of infectious individuals attain a given number (x – axis), which correlates with the time elapsed in the outbreak. $R_C$ is the observed maximum level of $R$ needed to result in a set-point of 100 or fewer infectious individuals after 60 days of confinement (y – axis). Although not shown, imposing mitigation measures in the range 1.0<$R_C$<2.5 prior to suppression does not change this basic result. Lines linking points aid visualization. See text for further details.

**Figure 1** shows that larger numbers of infectious individuals require stricter measures to attain the (arbitrary) set-point objective of 100 or fewer circulating infectious individuals after two months of measures. According to our simulations, a population the approximate size of Western European nations such as France, Spain, the UK and Germany would need to reduce transmission probabilities to near zero to attain the objective, should infectious numbers be on the order of 100,000. Early interventions in populations with 1 to 100 circulating infectious individuals would still require considerable measures, with reductions in baseline $R_0$=2.5 of 25% to 60%, respectively. Given cumulative case number doubling times observed in some countries of about 3 to 5 days (*17*) (and therefore *c.* 2 weeks between each log integer on the x-axis of **Figure 1**), the choice of measures without knowing their true impacts on the reproduction number could have resulted in insufficient flattening of the epidemic curve and valuable time lost. The data in **Table 2** support these insights, whereby suppression measures need to be increasingly strict in order to reduce the time necessary to obtain low target case numbers.

| **Duration (days)** | $R_C$=0.1<br>E,I | $R_C$=0.5<br>E,I | $R_C$=0.9<br>E,I |
|---|---|---|---|
| **0** | 11643,4645 | 11643,4645 | 11643,4645 |
| **10** | 2392,2261 | 5240,3643 | 9480,5475 |
| **20** | 492,500 | 2563,1801 | 8339,4829 |
| **30** | 102,105 | 1257,884 | 7340,4251 |
| **40** | 21,22 | 617,434 | 6460,3741 |
| **50** | 4,5 | 303,213 | 5684,3292 |
| **60** | 1,1 | 148,104 | 5000,2896 |

**Table 2.** Effect of suppression measure intensity and duration on the number of cases in incubation (E) and infectious (I) stages. N=70,420,854. $\mathcal{R}_0$=2.5. Start day of intervention=64. $E_{64}$=11,643, $I_{64}$=4,645.

We then asked how population size and the effectiveness of suppression measures would condition how subsequent mitigation measures attain objectives after 200 days. We conducted a 2 x 2 numerical experiment. The first variable was community size, taken either to be a small city of about 70,000 inhabitants or a medium-sized nation of about 70 million. (Additional numerical experiments not presented here indicate that the observed trends apply at least in the total population range of $10^4$ to $10^8$). The second variable explored was the effectiveness of previous suppression measures; we evaluated high effectiveness (a reset to a single infectious case) and a less, but still acceptable reset to 100 infectious cases. Clearly, any subsequent mitigation measures yielding $\mathcal{R}_C$<1.0 would result in infectious cases decreasing over time (and therefore be a successful outcome), but given the impacts of such measures on society, below we consider strategies that seek to contain a second epidemic by tuning $\mathcal{R}_C$ to between 1.0 and $\mathcal{R}_0$.

**Figure 2** shows how the effectiveness of suppression strategies and community size influence how subsequent mitigation measures affect epidemiological and clinical parameters. For example, regulating the infectious numbers to less than 10% of the total population requires $\mathcal{R}_C$ less than *c.*1.5, which is about a 40% reduction in the $\mathcal{R}_0$ assumed here (**Fig. 2A**). Peak levels of hospitalization can reach *c.*7%-10% should mitigation measures be 20% or less effective at reducing $\mathcal{R}_0$ (**Fig. 2B**). Such levels would exceed hospital bed capacity in most countries by at least an order of magnitude (*18*). Reducing peak hospitalization levels well below 1% (which is still too high for many health services) would require $\mathcal{R}_C$ close to 1.0. Finally, similar to the trends in **Figs. 2A,B**, fatalities are sensitive to the effectiveness of prior suppression measures and community size, indicating that $\mathcal{R}_C$ needs to be reduced towards unity for smaller communities and those unable to reduce infectious cases sufficiently during suppression measures (**Fig. 2C**). These results emphasize that epidemics could be contained by tuning $\mathcal{R}_C$ close to, but above 1.0, which would be more logistically and socially attainable than $\mathcal{R}_C$<1.0.

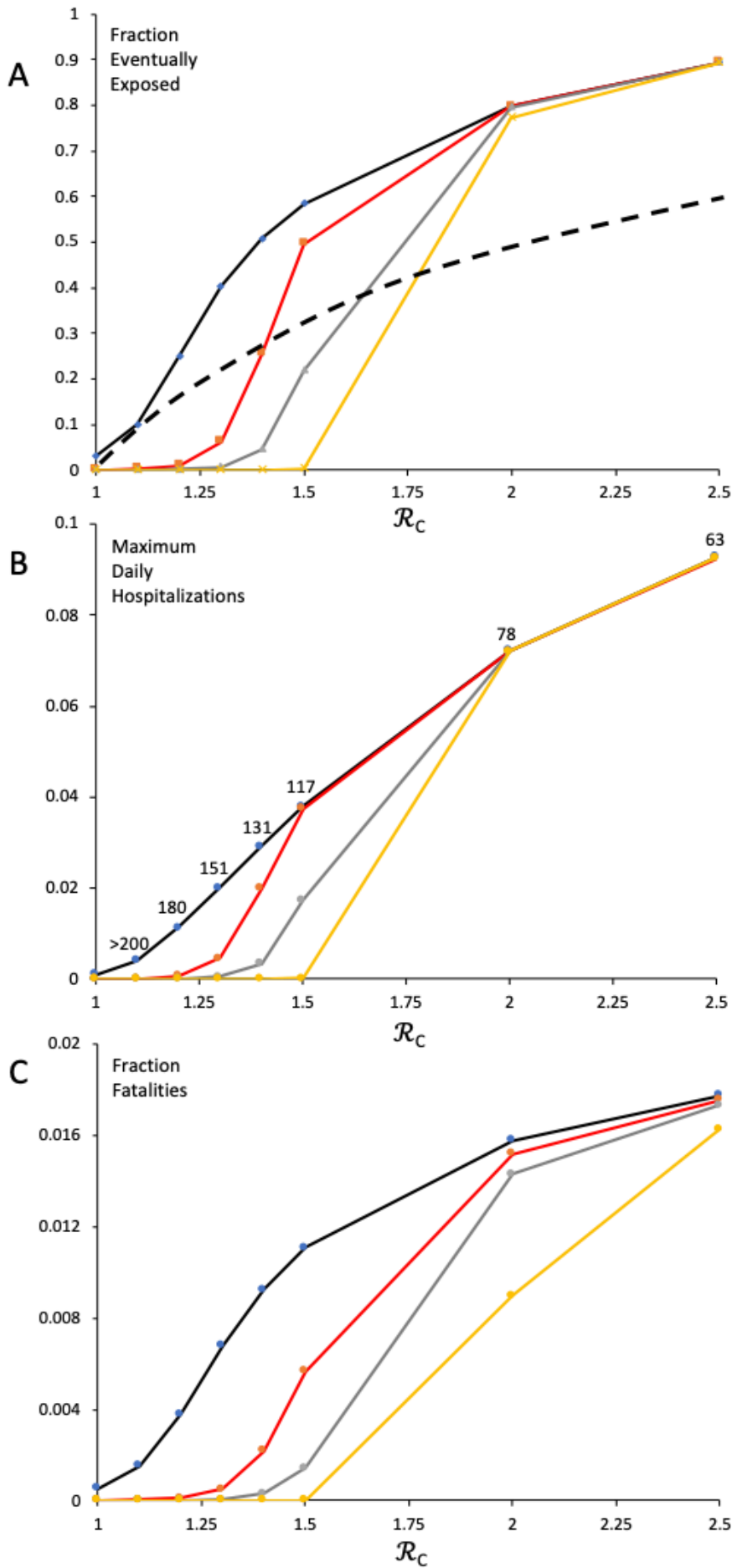

**Figure 2.** Effect of mitigation $\mathcal{R}_C$ on **(A)** The fraction of initially unexposed people who are eventually exposed $(E_t+I_t+R_t)$ at t=200 days after the measure starts; **(B)** The maximum daily fraction of the population needing hospitalization for any single day, up to 200 days after the measure starts; **(C)** The fraction of initially unexposed people who die during the 200 days of outbreak mitigation. Lines linking points aid visualization. Mitigation starts at: Yellow line, $I_0$=1, N=70 million; Gray line: $I_0$=100, N=70 million; Red line: $I_0$=1, N=70K; Black line: $I_0$=100, N=70K. $\mathcal{R}_0$=2.5.

Although not explored here, numerical simulations for parameter values associated with points above the dashed line in **(A)** were influenced by 'herd immunity', i.e., when the proportion of the population in the susceptible class, S/N<(1.0-1.0/$\mathcal{R}$) (*19*). Moreover, note that the result in **(A)** appears to contrast with the results in Steir et al. (*5*), who showed higher case growth rates with city size. The discrepancy can be explained by the different units employed in each study (case growth rate in (5) vs. fraction of total population infected at some point during a fixed time interval (this study)) and how $\mathcal{R}$ was estimated in (5) (found to be city-size dependent) vs. assumed invariant in the present study. Case growth rate (number of new cases on day t – number of new cases on day t-1 / number of new cases on day t-1) was found to increase with community size in the present study (not shown). Numbers above points in **(B)** refer to the day that maximum hospitalization occurs, and are only shown for the Black line conditions (note that when $\mathcal{R}_C$=1.0, maximum levels begin on day 90 and are constant thereafter; for $\mathcal{R}_C$=1.1, maximum levels occur after 200 days). See main text for additional details.

The herd immunity threshold is contingent on $\mathcal{R}_0$. But policies that reduce transmissions resulting in $\mathcal{R} < \mathcal{R}_0$ will also reduce the threshold. We observed that the attainment of herd immunity, whereby the rate of new infections is reduced below unity (when $S_t/N < 1.0 - 1.0/\mathcal{R}$), did not ensure the rapid end to an outbreak (**Fig. 2A**). Indeed, 'epidemic overshoot' (*20*) occurs in this simple model for all $\mathcal{R} > 1.0$. **Figure 3** extends the findings in **Fig. 2A** (limited to 200 days) to different points in the epidemic. We see that infected levels exceed the herd immunity threshold (black line) before maximum hospitalizations are observed (gray line). For $\mathcal{R}_C \geq 1.2$ the overshoot by the end of the epidemic (blue line) is between *c.*15-30% of the population. Moderate levels of mitigation can have major impacts in the short term (*cf.* $\mathcal{R}_C = 2.0$ vs. $\mathcal{R}_0 = 3.0$ at 60 days), but as the epidemic runs its course, the sensitivity of $\mathcal{R}_C$ exceeding 1.0 becomes apparent; for example, $\mathcal{R}_C = 1.2$ results in *c.*30% of the population infected at some point during the epidemic. We discuss the implications of this important finding for policies that 'leverage' immunity in the Conclusions.

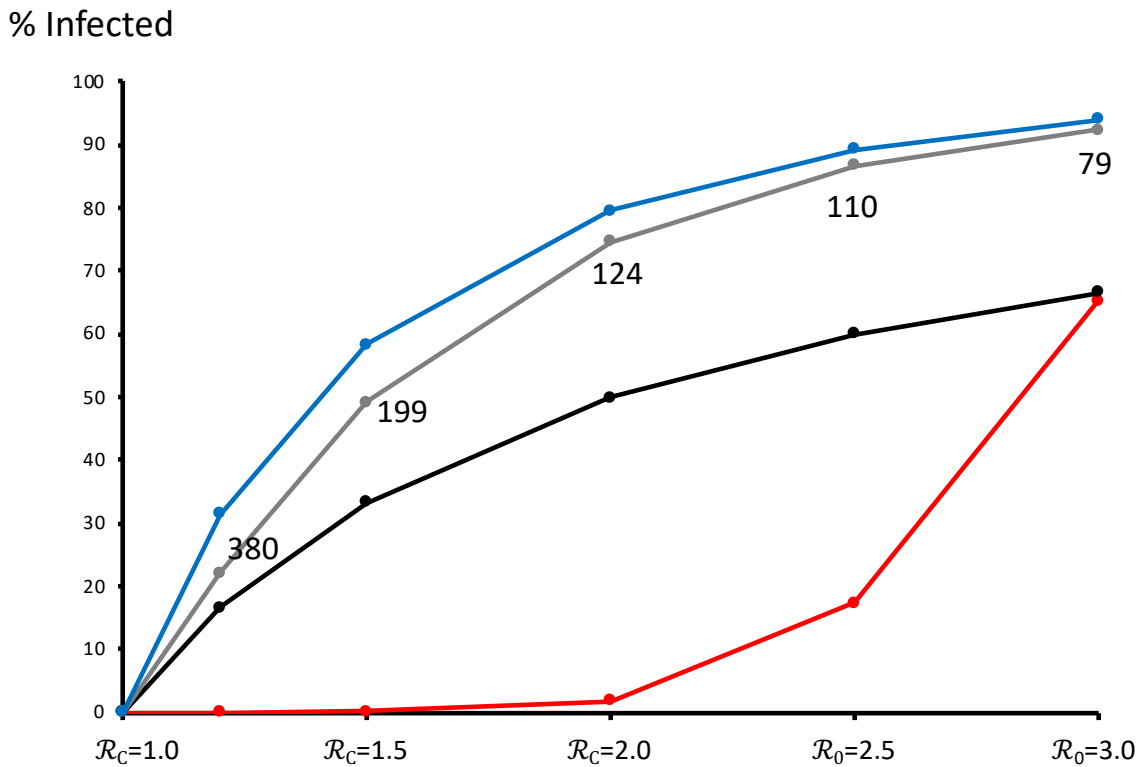

**Figure 3.** Epidemic overshoot: Effect of mitigation $\mathcal{R}_C$ on the percentage of an initially susceptible population infected by the virus at different points in the epidemic. $I_0=100$, N=70 million. Red line = 60 days after 100[th] infectious case; Black line = herd immunity threshold; Gray line = day of peak hospitalizations (numbers indicate day); Blue line = end of epidemic. Note as in Figure 2A, the herd immunity threshold is contingent on $\mathcal{R}$. $\mathcal{R}_0$ =2.5, 3.0 is shown for comparison. Lines linking points aid visualization. See main text for additional details.

Finally, whereas mitigation measures result in fewer cases (**Figs. 2, 3**), if $\mathcal{R}_C > 1.0$, then they also lengthen the course of the epidemic (**Figure 4**). An epidemic with no mitigation (example shown here is $\mathcal{R}_0 = 3.0$) is 95% complete 85 days after the 100[th] case. In contrast, an outbreak coming closest to the mitigation target of a flat epidemic ($\mathcal{R}_C$ =1.2) is 80% complete after 482 days and 95% at 585 days. The implications of long-term mitigation levels are discussed below.

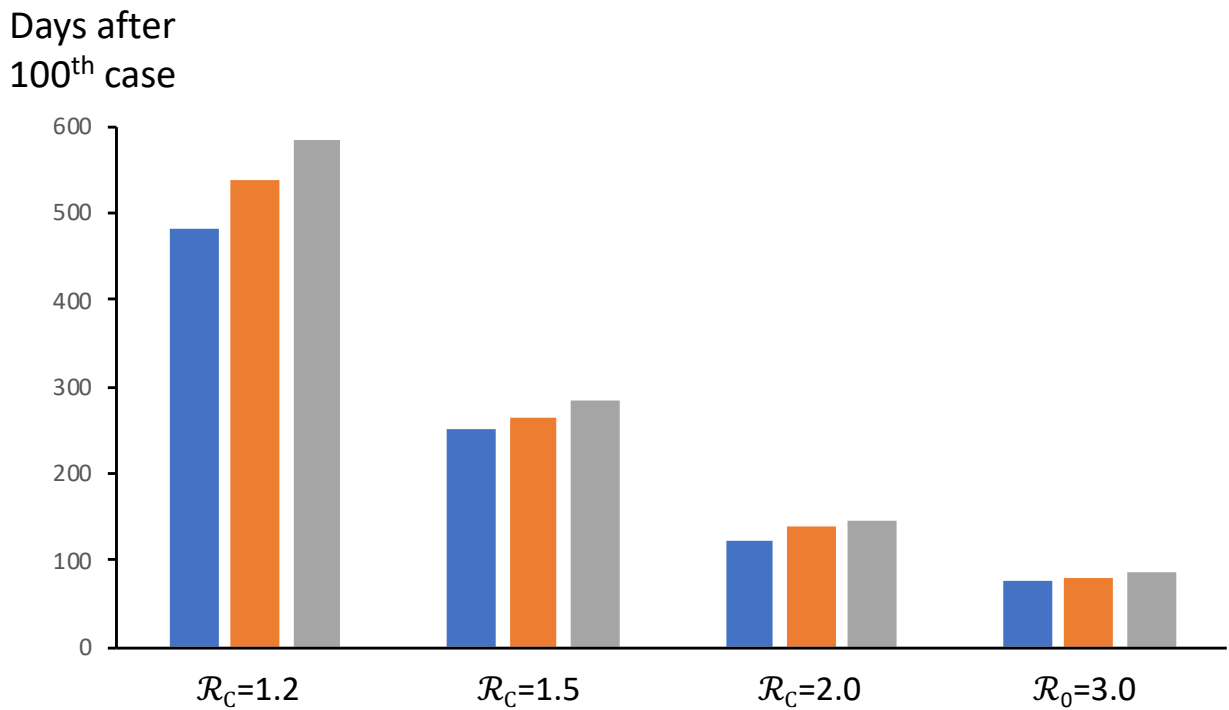

**Figure 4.** Effect of mitigation $\mathcal{R}_C$ on the number of days until 80%=blue, 90%=orange, or 95%=gray of eventual infections occur, during the full course of an epidemic. $I_0$=100, N=70 million. $\mathcal{R}_0$ =3.0 is shown for comparison See main text for additional details.

## Conclusions and future directions

The lockstep nature of COVID-19 outbreaks in different regions, countries and cities means that management practices in places further along the epidemic curve can inform those in earlier stages. Control strategies in one country, however, are not always applicable in others, due for example to cultural and logistical differences (*21*). Mathematical models based on empirical data have a role to play in adapting capacities to address epidemics, both as conceptual aids and management tools. The present study explored two generic types of intervention that can contribute to reducing the impact of COVID-19 epidemics. 'Suppression measures' would be adopted by communities that either initially decided not (or did, but were unable) to mitigate the exponential increase of new cases. Suppression reduces the reproduction number below 1.0, and in so doing lowers the number of infectious cases to a manageable level. 'Mitigation measures' may either be introduced as a preventive approach, whereby communities begin to manage very early in an outbreak, gradually introduced towards the end of a suppression strategy, or for countries unable or unwilling to enforce suppression measures, the only route possible to flatten the epidemic curve. The sequence of objectives observed in a number of countries, and likely to be seen in future months, can be symbolically represented as sequential stages:

$$\mathcal{R}_0 \rightarrow \mathcal{R}_C < \mathcal{R}_0 \rightarrow \mathcal{R}_C << 1.0 \rightarrow \mathcal{R}_C \approx 1.0 \rightarrow \mathcal{R}_{Ceff} < 1.0$$

The latter condition ($\mathcal{R}_{Ceff} < 1.0$) was not explored in this study, but 'leverages' any acquired immunity so as to lower the effective reproduction number $\mathcal{R}_{Ceff} \approx \mathcal{R}_C \, S/N$ below 1.0 (see discussion below).

The actual application of measures to meet this sequence of objectives is likely to be complex (*12-14*), and in particular, mitigation measures could go through multiple successive adjustments in the latter two phases represented above.

Our analysis presented here yield several important insights and predictions:

1. **The number of infectious cases at the start of suppression determine how effective these measures are at creating a reset**. Communities with more than 100,000 infectious cases essentially have to reduce transmission to zero in order to have about 100 cases after 60 days. Greater suppression measure stringency also means that with slightly longer confinement periods, an ever-lower set-point in the numbers of infectious individuals can be obtained. This permits reduced measure intensity in subsequent mitigations.

2. **The duration of suppression measures needed to attain an objective decreases with the intensity of measures**. Lowering $\mathcal{R}_C$ to or below 0.1 (i.e., a 96+% drop from $\mathcal{R}_0$=2.5) from just over 10,000 infectious cases would meet the objective of 100 after 30 days, and this approximately coincides with what was the Chinese data suggests (*1*). According to our simulations, an $\mathcal{R}_C$>0.5 would require more than 60 days to meet this same objective.

3. **Epidemic management is most effective if engaged when infectious cases are low enough to preserve health service capacities**. As above, this is best done either near the start of an outbreak or once a suppression reset has obtained. Although in this scenario, $\mathcal{R}_C$<1.0 is a sufficient condition for continued suppression, tactics that give weight to individual freedoms, yet keep $\mathcal{R}_C$ above, but sufficiently close to 1.0 can limit epidemic consequences (see also below), including morbidities and mortalities and the saturation of health service capacities. Data analysis of 11 European countries (*14*) supports the basic quantitative predictions set out in the present study, and similar analyses of richer, near-future data sets will be needed to refine target parameter values so as to yield regional epidemic and global pandemic management objectives.

4. **Epidemics do not suddenly stop when the fraction of susceptible individuals drops below the herd immunity threshold (*20,22*).** We find that although herd immunity is predicted to occur when 60 to 70% of the population have been infected and become immune (based on parameters in **Table 1**), if an outbreak is left unchecked, then the virus will eventually infect approximately 90% of the population ('epidemic overshoot'). According to our model, some level of overshoot always occurs for mitigation measures, but the degree depends on the intensity of such members. For example, measures that lower $\mathcal{R}_0$ by 50% from 3.0 to 1.5 will still result in approximately 60% of the population eventually becoming infected, even though the herd immunity threshold is about 33%.

5. **Mitigation extends the course of an outbreak (Figure 4)**. Mitigation, even if less restrictive compared to suppression, has impacts on health services, society and the economy. Maintaining $\mathcal{R}_C \geq 1.0$ is therefore a short-term solution to contain an epidemic whilst preserving certain liberties. Longer-term (months) measures will need to lower the effective reproduction number $\mathcal{R}_{Ceff} \approx \mathcal{R}_C \, S/N$ below 1.0, so as to attenuate and finally stop the outbreak.

This latter prediction highlights the potential for any acquired immunity to be 'leveraged' so as to ease-off on restrictions (thereby *increasing* $\mathcal{R}_C$), but nevertheless effectively mitigate or even suppress the outbreak. For example, if 15% of a population were previously exposed and immune, then this suggests that $\mathcal{R}_C < 1.18$ (equivalently $\mathcal{R}_{Ceff} < 1.0$) would suppress the outbreak. This 18% increase would enable less-intense measures compared to the situation at the start of an outbreak, where $\mathcal{R}_C < 1.0$ is necessary. Although more study is needed, we suggest that this strategy could be progressive, meaning that as fractions of the population with acquired immunity continue to increase, so too could the easing-off that results in suppression (i.e., $\mathcal{R}_{Ceff} < 1.0$ is maintained, despite purposeful increases in $\mathcal{R}_C$). A similar strategy called 'shield immunity' has recently been proposed by Weitz and coworkers (*23*), and consists of the intentional deployment of immune individuals so as to reduce transmission and outbreaks. The potential use of such leveraging will only become clear once long-lasting immunity is better understood.

Of the many limitations to our analysis, two in particular merit further study. First, the model assumes random mixing of individuals. Real infection networks are far more complex, and may involve (*i*) significant spatial structuring, (*ii*) different numbers of contacts per individual and through time and travel, (*iii*) epidemiological class effects (such as age, quarantined, hospitalized), and (iv) persistence of the virus in the external environment. Mathematical models incorporating heterogeneous contact structures (e.g., *5, 7, 24*) will have a role to play in indicating the effectiveness of different control measures. Analyses similar to (*14*) should explore both realistic contact structures and community-specific values of epidemic and health service parameters.

Second, future analyses will need to translate actual tactics into their effects on different epidemiological parameters, and specifically $\mathcal{R}_C$. A number of tactics have been proposed and some variously adopted by different communities, including: spatial distancing, quarantining, hand washing, wearing masks, gloves; diagnostics such as contact tracing, and virus and antibody testing; and interventions such as employing repurposed drugs and developing vaccines. Such approaches will need to be employed in complementary ways, since no single one is likely to attain reset or management objectives. Moreover, calibration of $\mathcal{R}_C$ in particular will require accurate assessments of the contribution of asymptomatic transmission to the propagation of the virus, the latter having been recently demonstrated in mathematical models to potentially influence COVID-19 dynamics (*25*).

In conclusion, the simple model analyzed here is not an instrument to develop precise, actionable strategies. Rather, it is a conceptual tool that identifies some of the important parameters, and generates testable hypotheses of how these could affect outbreak management outcomes.

## Acknowledgements

I thank the members of the EEC research team at the Institute for Evolutionary Sciences, University of Montpellier, Carl Bergstrom, Katrina Lythgoe, Andrew Dobson, and Joshua Weitz for comments and discussions.